\documentclass[epjST]{svjour}
\usepackage{graphicx,subfigure,amssymb}
\usepackage{epstopdf}
\usepackage{color}
\begin{document}
\title{Kalman-Takens filtering in the presence\\ of dynamical noise }
\author{Franz Hamilton\inst{1} \and Tyrus Berry\inst{2} \and Timothy Sauer\inst{2}\fnmsep\thanks{\email{tsauer@gmu.edu}} }
\institute{North Carolina State University, Raleigh, NC 27695, USA \and George Mason University, Fairfax, VA 22030, USA }
\abstract{
The use of data assimilation for the merging of observed data with dynamical models is becoming standard in modern physics. If a parametric model is known, methods such as Kalman filtering have been developed for this purpose. If no model is known, a hybrid Kalman-Takens method has been recently introduced, in order to exploit the advantages of optimal filtering in a nonparametric setting.   This procedure replaces the parametric model with dynamics reconstructed from delay coordinates, while using the Kalman update formulation to assimilate new observations. We find that this hybrid approach results in comparable efficiency to parametric methods in identifying underlying dynamics, even in the presence of dynamical noise.  By combining the Kalman-Takens method with an adaptive filtering procedure we are able to estimate the statistics of the observational and dynamical noise.  This solves a long standing problem of separating dynamical and observational noise in time series data, which is especially challenging when no dynamical model is specified.
} %end of abstract
\maketitle
\section{Introduction}
\label{intro}
Methods of data assimilation are heavily used in geophysics and have become common throughout physics and other sciences.  When a parametric, physically motivated model is available, noise filtering and forecasting in a variety of applications are possible. Although the original Kalman filter \cite{kalman} applies to linear systems, 
more recent approaches to filtering such as the Extended Kalman Filter (EKF) and Ensemble Kalman Filter (EnKF) \cite{kalnay,evensen,julier1,julier2,schiff} allow forecasting models to use the nonlinear model equations to compute predictions that are close to optimal. 

In some cases, a model is not be available, and in other cases, all available models may be inaccurate.  In geophysical processes, basic principles may constrain a variable as a function of other driving variables in a known manner, but the driving variables may be unmodeled or modeled with large error \cite{reichle05,hersbach07,arnold13,BerryHarlim14}. Moreover, in numerical weather prediction codes, physics on the large scale is typically sparsely modeled \cite{Lorenz96,Palmer01}. Some recent work has considered the case where only a partial model is known \cite{Hamilton15,BerryHarlim15}.

Under circumstances in which models are not available, Takens' method of attractor reconstruction \cite{Takens,Packard80,Embed,Sauer04} has been used to reconstruct physical attractors from data. 
The dynamical attractor is typically represented by vectors of  delay coordinates constructed from time series observations, and approaches to prediction, control, and other time series applications  have been proposed \cite{Coping94,Abarbanel96,Kantz04}. In particular, time series prediction algorithms locate the current position in the delay coordinate representation and use analogues from previously recorded data to establish a  predictive statistical model \cite{Farmer87,Casdagli89,SugiharaMay90,Lenny92,Jimenez92,Sauer94,Sugihara94,Schroer98,Kugiumtzis98,Yuan04,Hsieh05,Strelloff06,Regonda05,Schelter06}. 
However, Takens' method is designed (and proved to work) for strictly deterministic systems, and the effects of noise, both dynamic and observational, is not thoroughly understood.

Recently, a method was introduced that would merge Takens' nonparametric attractor reconstruction approach with Kalman filtering. 
Since the model equations governing the evolution of the system are unknown, the dynamics are reconstructed nonparametrically using delay-coordinate vectors, and used to replace the model. The Kalman-Takens algorithm \cite{PRX} is able to filter noisy data with comparable performance to parametric filtering methods that have full access to the exact model. The fidelity of this algorithm follows from the fact that Takens' theorem \cite{Takens} states roughly that in the large data limit, the equations can be replaced by the data.  The surprising fact, shown in \cite{PRX}, is that this fidelity is robust to observational noise which invalidates the basic theory of \cite{Takens}, although a much more complex theory developed in \cite{BroomheadStark} suggests such a robustness.  In fact, by implementing the Kalman update, it is suggested in \cite{PRX}  that the nonparametric representation of the dynamics is able to handle substantial observational noise in the data.

In this article we study the effects of dynamic noise (also known as system noise or process noise) on the Kalman-Takens algorithm. We will verify the effectiveness of the method on stochastic differential equations with significant levels of system noise, and show that nonparametric prediction can be improved by the filter almost to the extent of matching the performance of the exact parametric model. In section 2 we present the specifics of the Kalman-Takens method. A key to effective application of any Kalman based algorithm is knowing the covariance matrices of the system and observation noise, which are particularly difficult to separate when no model is known. In section 3 we present an adaptive method for inferring statistics of the system and observational noise as part of the filtering procedure. The adaptive filtering method naturally complements the Kalman-Takens method by implicitly fitting a fully nonparametric stochastic system to the data.  These methods are combined in section 4, where applications to dynamical data with variable settings of delay coordinates are explored.

\section{Kalman-Takens filter}
\label{sec:2}

We recall the standard notion of the Kalman filter in the case where the model $f$ and observation function $g$ are known. Consider a nonlinear stochastic system with $n$-dimensional state vector $x$ and $m$-dimensional observation vector $y$ defined by 
\begin{eqnarray} \label{e1}
x_{k+1} &=& f(x_{k},t_{k})+w_{k}\\
y_{k} &=& g(x_{k},t_{k})+v_{k}\nonumber
\end{eqnarray}
where $w_{k}$ and $v_{k}$ are white noise processes with covariance matrices $Q$ and $R$, respectively. We begin by describing the filtering procedure in the case when $f$ and $g$ are known.

We are chiefly interested in nonlinear systems, so we will describe a version of Kalman filtering that is common in this case. The ensemble Kalman filter (EnKF) \cite{kalnay} represents a nonlinear system at a given instant as an ensemble of states.  Here we initialize the filter with state vector $x^+_0 = 0_{n\times 1}$ and covariance matrix $P^+_0 = I_{n\times n}$. At the $k$th step, the filter produces an estimate of the state $x^+_{k-1}$ and the covariance matrix $P^+_{k-1}$, which estimates the covariance of the error between the estimate $x^+_{k-1}$ and the true state. In the unscented version of EnKF \cite{Simon}, the singular value decomposition is used to find the symmetric positive definite square root $S^+_{k-1}$ of the matrix $P^+_{k-1}$, allowing us to form an ensemble of $E$ state vectors, where the $i^{th}$ ensemble member is denoted $x_{i,k-1}^+$. The ensemble vectors $x_{i,k-1}^+$ are formed by adding and subtracting the columns of $S^+_{k-1}$ to the estimate $x^+_{k-1}$ to produce and ensemble with mean $x^+_{k-1}$ and covariance $P^+_{k-1}$ \cite{Simon,julier1,julier2}. In other words, the ensemble statistics match the current filter estimates of the mean and covariance. The ensemble can also be rescaled by introducing weights for each ensemble member as described in \cite{julier1,julier2}.

The model $f$ is applied to the ensemble, advancing it one time step, and then observed with function $g$. The mean of the resulting state ensemble gives the prior state estimate $x_k^-$ and the mean of the observed ensemble is the predicted observation $y_k^-$.  Denoting the covariance matrices $P_k^-$ and $P_k^y$ of the resulting state and observed ensemble, and the cross-covariance matrix $P_k^{xy}$ between the state and observed ensembles, we define
\begin{eqnarray}\label{e2}
P_k^- &=& \sum_{i= 1}^{E} \left(x_{ik}^-  -x_k^-\right) \left(x_{ik}^- -x_k^-\right)^T + Q\nonumber\\
P_k^y &=&  \sum_{i= 1}^{E} \left(y_{ik}^-  -y_k^-  \right) \left(y_{ik}^-  -y_k^- \right)^T + R \nonumber\\
P_k^{xy} &=& \sum_{i= 1}^{E} \left(x_{ik}^-  -x_k^- \right) \left(y_{ik}^-  -y_k^-  \right) ^T.
\end{eqnarray}
Given the observation $y_k$, the equations
\begin{eqnarray} \label{e3}
K_k &=& P^{xy}_k(P^{y}_k)^{-1}\nonumber\\
P^{+}_k &=& P^{-}_k-P^{xy}_k(P^{y}_k)^{-1}P^{yx}_k\nonumber\\
{x}^{+}_k &=& {x}^{-}_k+K_k\left(y_k-{y}_k^- \right).
\end{eqnarray}
 update the state and covariance estimates. We will refer to this as the parametric filter, since a full set of model equations are assumed to be known. The matrices $Q$ and $R$ are parameters which represent the covariance matrices of the system noise and the observation noise respectively.  Often the true statistics of these noise processes are unknown, and examples in \cite{qr13} have shown that accurate estimates of $Q$ and $R$ are crucial to obtaining a good estimate of the state. In section 3, we describe an algorithm developed in \cite{qr13} for adaptive estimation of $Q$ and $R$ which was developed for the case when the dynamical model $f$ and observation function $g$ are known.

Contrary to (\ref{e1}), our assumption in this article is that neither the model $f$ or observation function $g$ are known, making outright implementation of the EnKF impossible. Instead, the filter described here requires no model while still leveraging the Kalman update described in (\ref{e3}). The idea is to replace the system evolution, traditionally done through application of $f$, with advancement of dynamics nonparametrically using delay-coordinate vectors. We describe this method with the assumption that a single variable is observable, say $y$, but the algorithm can be easily generalized to multiple system observables. In addition, we will assume in our examples that noise covariances $Q$ and $R$ are unknown and will be updated adaptively as in \cite{qr13}.

The idea behind Takens' method is that given the observable $y_k$, the delay-coordinate vector 
\[ x_k = [y_k, y_{k-1}, \ldots, y_{k-d}] \] 
accurately captures the \emph{state} of the dynamical system, where $d$ is the number of delays.  In the case where the noise variables are removed from (\ref{e1}), the vector $x_k$ provably represents the state of the system for $d+1 > 2n$ as shown in \cite{Takens,Embed} (note that the embedding dimension is $d+1$ since $d$ delays are added to the original state). An extension of this result to the stochastic system (\ref{e1}) can be found in \cite{BroomheadStark}. Current applications of Takens delay-coordinate reconstruction have been restricted to forecasting; typically by finding historical delay-coordinate states which are close to the current delay-coordinate state and interpolating these historical trajectories. A significant challenge in this method is finding `good' neighbors, which accurately mirror the current state, especially in real applications when both the current and historical states are corrupted by observational noise.  The goal of the Kalman-Takens filter is to quantify the uncertainty in the state and reduce the noise using the Kalman filter.

In order to integrate the Takens reconstruction into the Kalman filter, at each step of the filter an ensemble of delay vectors is formed. The advancement of this ensemble forward in time requires a nonparametric technique to serve as a local proxy $\tilde{f}$ for the unknown model $f$.  Given a specific delay coordinate vector $x_k=[y_k, y_{k-1}, \ldots, y_{k-d}],$ we locate its $N$ nearest neighbors (with respect to Euclidean distance) %$[y_{k'}, y_{k'-1}, \ldots, y_{k'-d}]$, $[y_{k''}, y_{k''-1}, \ldots, y_{k''-d}], \ldots,  [y_{k^N}, y_{k^N-1}, \ldots, y_{k^N-d}]$ 
$x_{i_1},...,x_{i_N}$ where
\[ x_{i_j} = [y_{i_j}, y_{i_j-1}, \ldots, y_{i_j-d}] \]
are found from the noisy training data. Once the neighbors are found, the known one step forecast values $y_{i_1+1},y_{i_2+1},...,y_{i_N+1}$%$y_{k'+1}, y_{k''+1}, \ldots, y_{k^N+1}$ values 
 are used with a local model to predict $y_{k+1}$. In this article, we use a locally constant model which in its most basic form is an average of the nearest neighbors, namely
\begin{eqnarray*}
\tilde{f}(x_k)=\left[\frac{y_{i_1+1} + y_{i_2+1}+ \ldots+ y_{i_N+1}}{N},y_k, \ldots, y_{k-d+1}\right].
%\tilde{f}(x_k)=\left[\frac{y_{k'+1} + y_{k''+1}+ \ldots+ y_{k^N+1}}{N},y_k, \ldots, y_{k-d+1}\right].
\end{eqnarray*}

This prediction can be further refined by considering a weighted average of the nearest neighbors where the weights are assigned as a function of the neighbor's distance to the current delay-coordinate vector. Throughout the following examples 20 neighbors were used. This process is repeated to compute the one step forecast $\tilde f$ applied to each member of the ensemble. Once the full ensemble has been advanced forward in time by $\tilde f$, the remaining EnKF algorithm is then applied using equations (\ref{e2}) as described above, and our delay-coordinate state vector is updated according to (\ref{e3}). This method was called the {\it Kalman-Takens filter} in \cite{PRX}.

\begin{figure}[bt]
\begin{center}
\subfigure[]{\includegraphics[width=.48\textwidth]{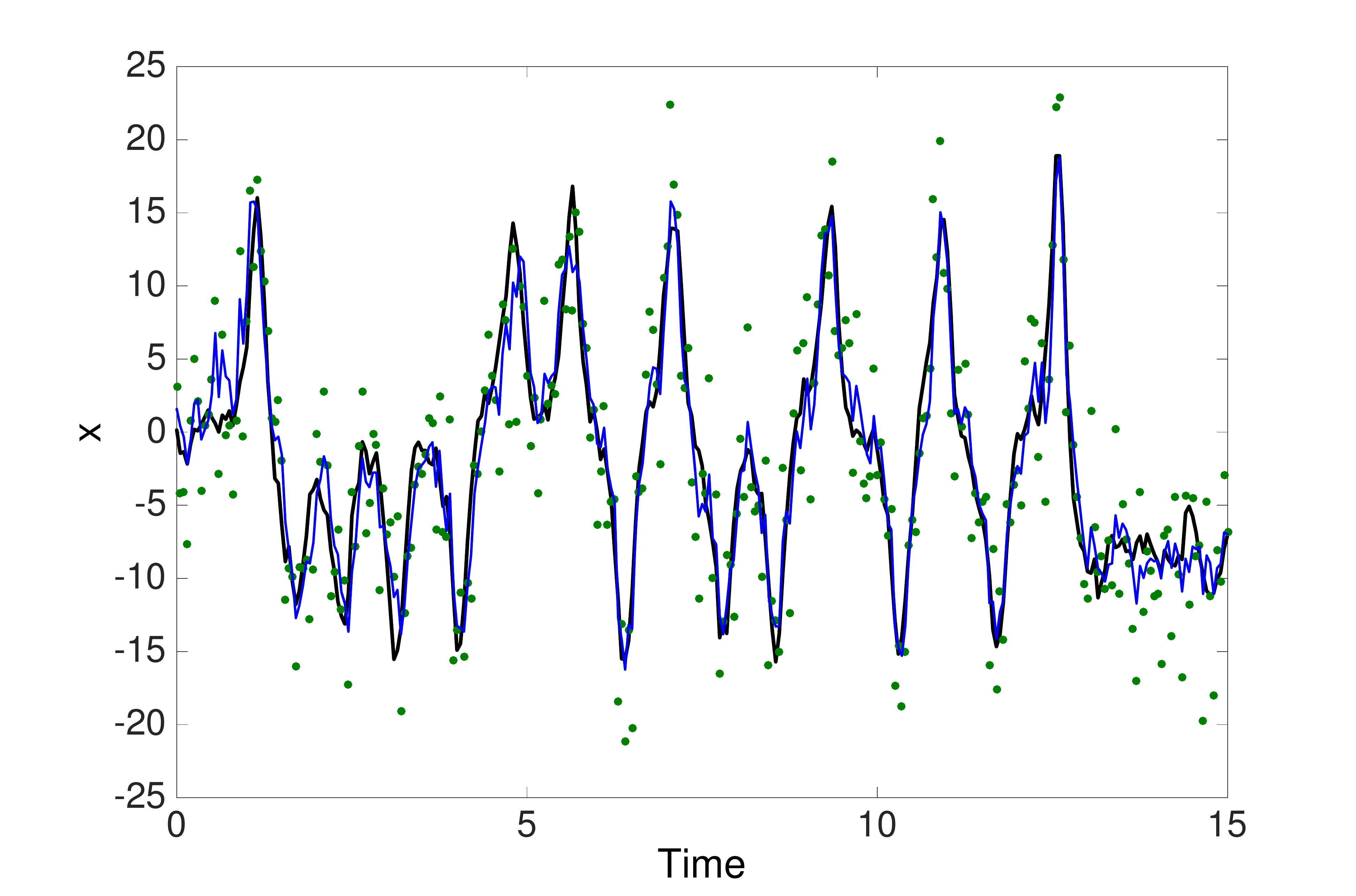}}
\subfigure[]{\includegraphics[width=.48\textwidth]{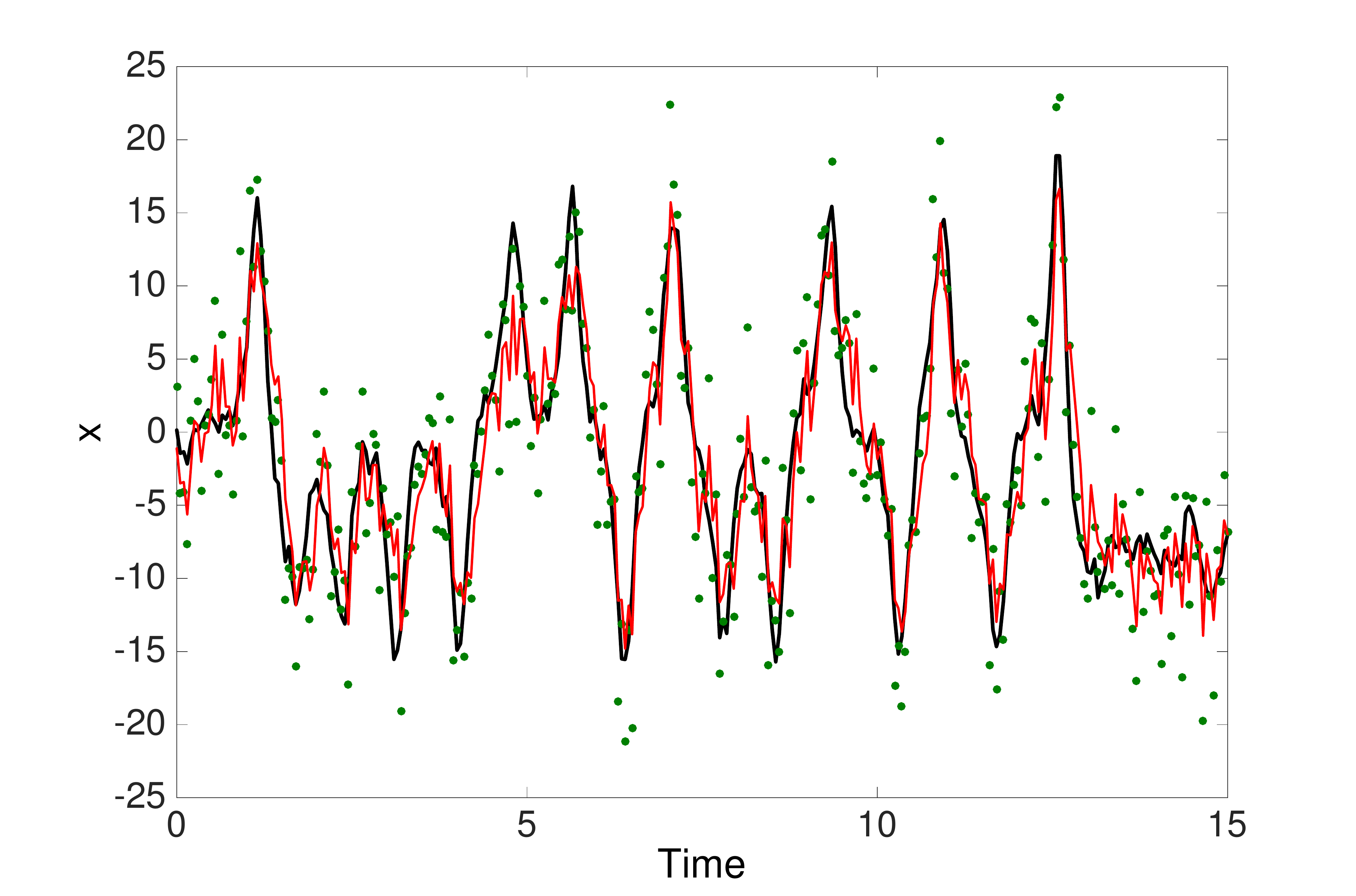}}
\end{center}
\caption{Filtering comparison of the Lorenz-63 $x$ variable when the system is affected by Gaussian dynamical noise with total variance of 2.4 (black solid line). Observations (green circles) of the stochastic signal perturbed by Gaussian observational noise with variance of 20 (signal RMSE of 4.49). (a) Parametric filter reconstruction (solid blue line) results in an RMSE of 2.34. (b) Kalman-Takens filter with 2 delays (solid red curve) results in RMSE of 2.95. }
\label{fig:1}       % Give a unique label
\end{figure}

As an example we consider the following SDE based on the Lorenz-63 system \cite{Lorenz63}
\begin{eqnarray} \label{e4}
\dot{x} &=& \sigma(y-x)+ \xi \dot{W}_x\nonumber\\
\dot{y} &=& x(\rho-z)-y + \xi \dot{W}_y\\
\dot{z} &=& xy-\beta z + \xi \dot{W}_z\nonumber
\end{eqnarray}
where $\sigma = 10$, $\rho = 28$, $\beta = 8/3$ and $\dot{W}$ is white noise with unit variance. Assume we have a noisy set of training data points
\[ y(t_k) = x(t_k) + \eta_k \]
where $k=1,2,...,M$ and $y_k = y(t_k)$ is a direct observation of the $x$ variable corrupted by independent Gaussian perturbations $\eta_k$ with mean zero and variance $R^o$. Using this data, we want to develop a nonparametric forecasting method to predict future $x$ values of the system. However, due to the presence of the noise $\eta_k$, outright application of a prediction method leads to inaccurate forecasts. 

If knowledge of (\ref{e4}) were available, the standard parametric filter could be used to assimilate the noisy $x$ observations to the model, generate a denoised estimate of the $x$ variable, and simultaneously estimate the unobserved $y$ and $z$ variables. This denoised state estimate could then be forecast forward in time using (\ref{e4}), and we refer to this process and the parametric forecast.

In contrast, the Kalman-Takens method assumes no knowledge of the underlying dynamics (\ref{e4}) or even the observation function.  Instead, the Kalman-Takens method works directly with the noise observations $y_k$ and defines a proxy for the underlying state by the delay vector $[y_k, y_{k-1}, \ldots, y_{k-d}]$.  By applying the Kalman-Takens method to the entire training data set, we reduce the observation noise in the training data which will improve future filtering and also provide better neighbors to improve forecasting.  

\begin{figure}[bt]
\begin{center}
\subfigure[]{\includegraphics[width=.48\textwidth]{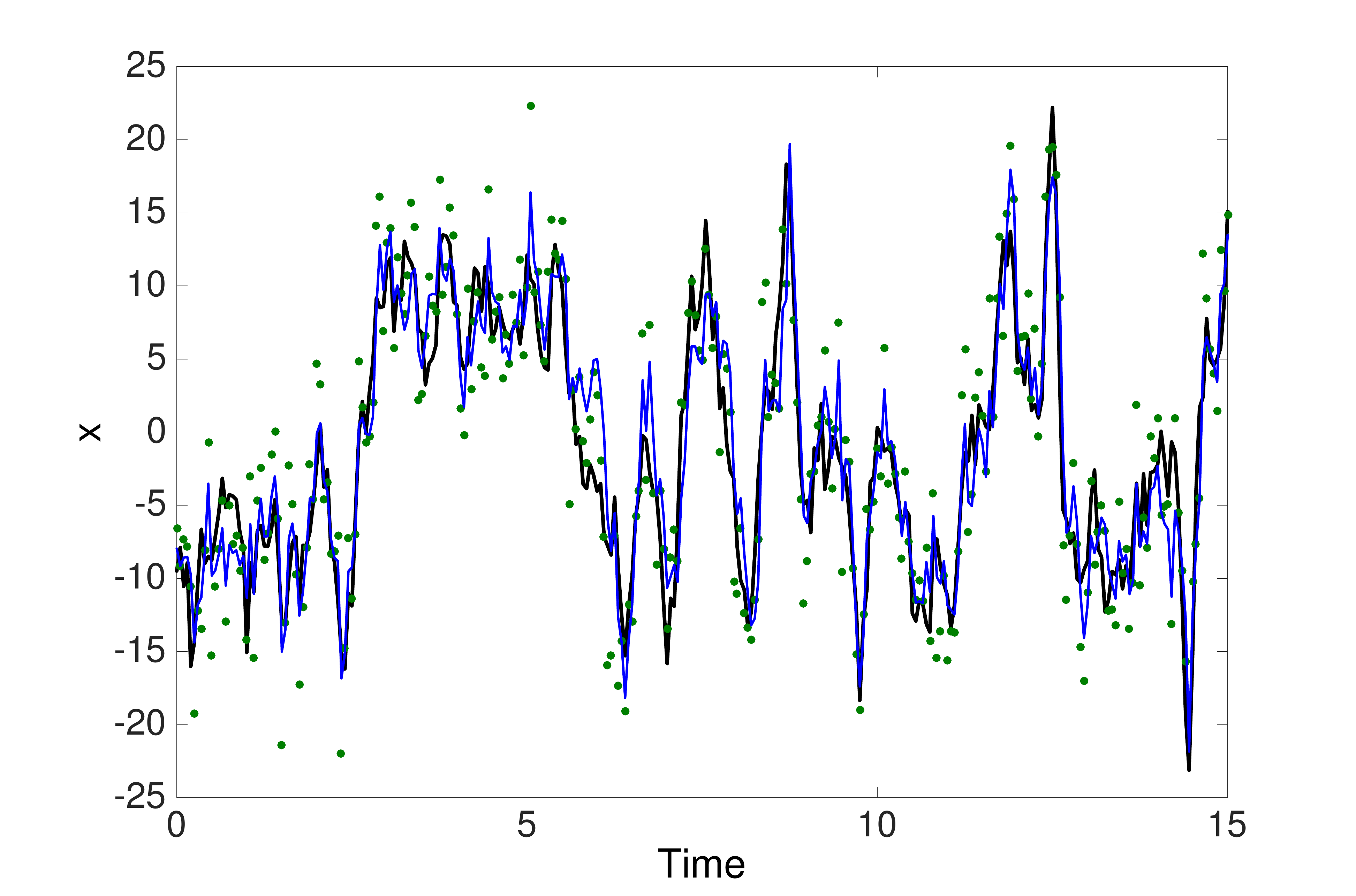}}
\subfigure[]{\includegraphics[width=.48\textwidth]{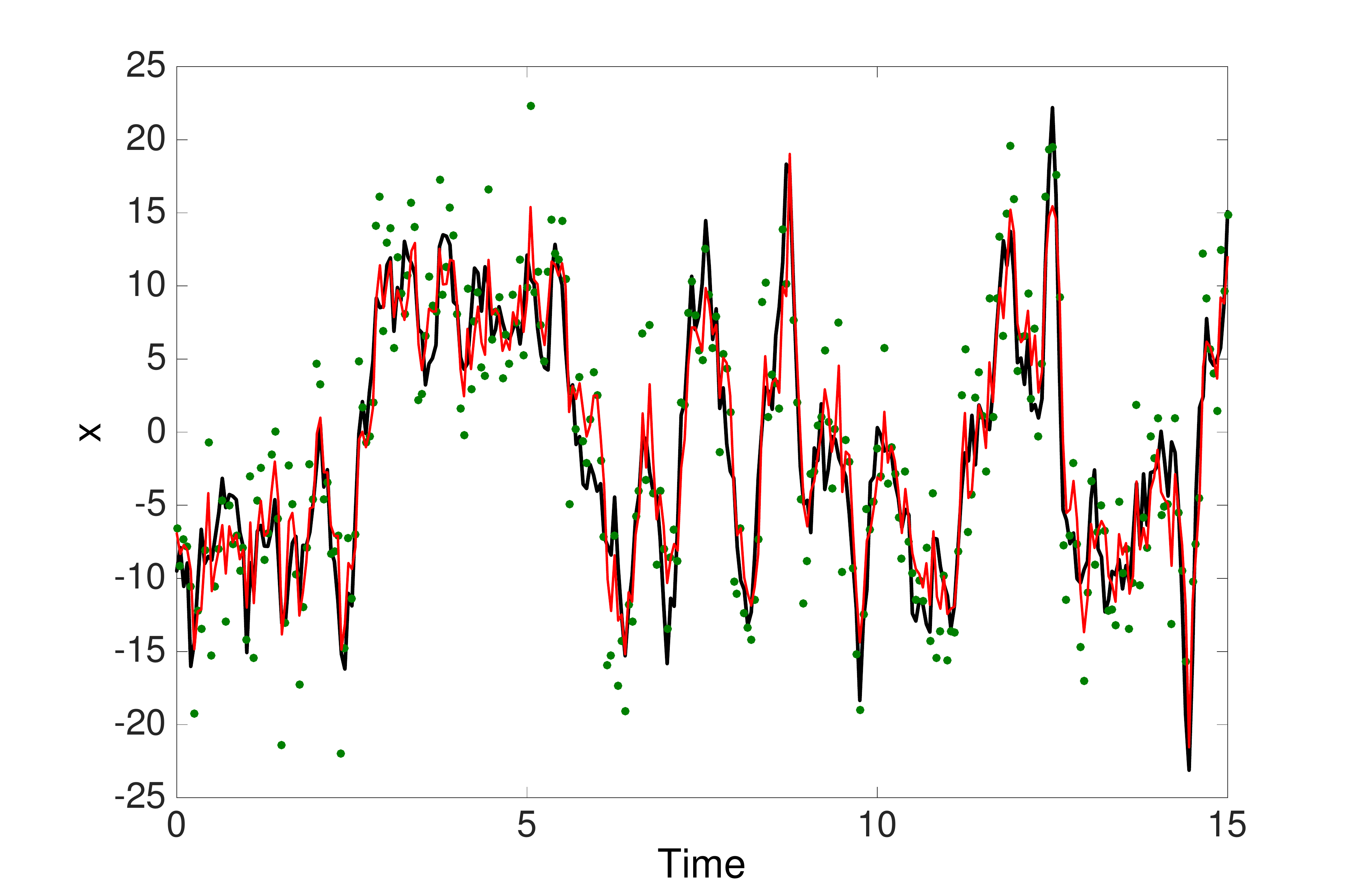}}
\end{center}
\caption{Filtering comparison of the Lorenz-63 $x$ variable when the system is affected by Gaussian dynamical noise with total variance of 15 (black solid line). Observations (green circles) of the stochastic signal perturbed by Gaussian observational noise with variance of 20 (signal RMSE of 4.49). (a) Parametric filter reconstruction (solid blue line) results in an RMSE of 3.21. (b) Kalman-Takens filter with 2 delays (solid red curve) results in RMSE of 3.29.}
\label{fig:2}       % Give a unique label
\end{figure}

Fig.~\ref{fig:1} shows a comparison of ensemble Kalman filtering with and without the model. Fig.~\ref{fig:1}(a) shows the standard parametric EnKF applied to the $x$-coordinate of the Lorenz SDE (\ref{e4}) with system noise variance $\xi^2=0.8$ (the total variance of the system noise is 2.4 since there are three independent noise variables), and observational noise variance $R^o=20$. Fig.~\ref{fig:1}(b) shows the Kalman-Takens filter applied to the same data. Compared to the SDE solution $x(t_k)$ without observational noise (black curve), the parametric filter fits slightly better, but the Kalman-Takens filter does almost as well without knowing the model equations. A training set of $M=8000$ points were used. Fig.~\ref{fig:2} shows the same comparison with a higher level of system noise $\xi^2=5$ (total system noise variance is 15).

\section{Adaptive estimation of $Q$ and $R$}
\label{sec:3}

A persistent problem in time series analysis is distinguishing system noise from observational noise.  This is a particularly important issue since observational noise distorts the forecast, and needs to be removed, whereas the system noise affects the future of the system and thus should not be removed.  In other words, the state estimate that will give the optimal forecast includes the system noise perturbation, but does not include any of the observation noise perturbations.  While obtaining this perfect estimate is typically impossible, getting as close as possible requires knowing the statistics and correlations of the two noise processes.  This is captured in the structure of the Kalman equations (\ref{e2}) and (\ref{e3}) which gives the provably optimal estimate of the state for linear systems with additive Gaussian system and observation noise.  The presence of the noise covariance matrices $Q$ and $R$ in the Kalman equations shows how these parameters determine the optimal estimate.    

Since we cannot assume that the noise covariance matrices $Q$ and $R$ are known, we used a recently-developed method  \cite{qr13} for the adaptive fitting of these matrices as part of the filtering algorithm. The key difficulty in estimating these covariances is disambiguating the two noise sources.  The method of \cite{qr13} is based on the fact that system noise perturbations affect the state at future times, whereas observation noise perturbations only affect the current time.  

The method uses the innovations $\epsilon_k\equiv y_k-y_k^-$ in observation space from (\ref{e2}) to update the estimates $Q_k$ and $R_k$ of the covariances $Q$ and $R$, respectively, at step $k$ of the filter.  In order to estimate these two quantities, we will compute two statistics of the innovations.  The first statistic is the outer product $\epsilon_k \epsilon_k^\top$ and the second statistic is the lagged outer product $\epsilon_k \epsilon_{k-1}^\top$. Intuitively, the first statistic only includes information about the observation noise covariance, whereas the second statistic contains information about the system noise covariance.

We produce empirical estimates $Q^e_{k-1}$ and $R^e_{k-1}$ of $Q$ and $R$ based on the innovations at time $k$ and $k-1$ using the formulas
%\begin{align}\label{qrestimates}
%P^e_k &= (H_{k+1}F_k)^{-1}\left(\epsilon_{k+1}\epsilon_k^T + H_kF_kK_k\epsilon_k\epsilon_k^T \right)H_k^{-T} \nonumber \\
%Q^e_k &=  P^e_k - F_{k-1}P^a_{k-1}F_{k-1}^T \nonumber\\
%R^e_k &= \epsilon_k\epsilon_k^T - H_kP^f_kH_k^T
%\end{align}
\begin{eqnarray}\label{qrestimates}
P_{k-1}^e &= &F_{k-1}^{-1}H_k^{-1}\epsilon_k\epsilon_{k-1}^T H_{k-1}^{-T} + K_{k-1}\epsilon_{k-1}\epsilon_{k-1}^T H_{k-1}^{-T}  \nonumber \\
Q_{k-1}^e &=&  P_{k-1}^e - F_{k-2}P_{k-2}^aF_{k-2}^T \nonumber \\
R_{k-1}^e &= &\epsilon_{k-1}\epsilon_{k-1}^T - H_{k-1}P_{k-1}^fH_{k-1}^T.
\end{eqnarray}
It was shown in \cite{qr13} that $P^e_{k-1}$ is an empirical estimate of the background covariance at time index $k-1$.  Notice that (\ref{qrestimates}) requires local linearizations $F_{k-1}$ of the dynamics and $H_{k-1}$ of the observation function.  While there are many methods of finding these linearizations, we use the method introduced in \cite{qr13} and used in \cite{PRX} which is based on a linear regression.  In particular, to determine $F_{k-1}$ a linear regression is applied between the ensemble before and after the dynamics $\tilde f$ are applied.  Similarly, to determine $H_{k-1}$ a linear regression is applied between the ensemble before and after the observation function is applied. Notice that this procedure requires us to save the linearizations $F_{k-2},F_{k-1}, H_{k-1}, H_k$, innovations $\epsilon_{k-1}, \epsilon_k$, and the analysis $P_{k-2}^a$ and Kalman gain matrix, $K_{k-1}$, from the $k-1$ and $k-2$ steps of the EnKF.

\begin{figure}
\begin{center}
\subfigure[]{\includegraphics[width=.48\textwidth]{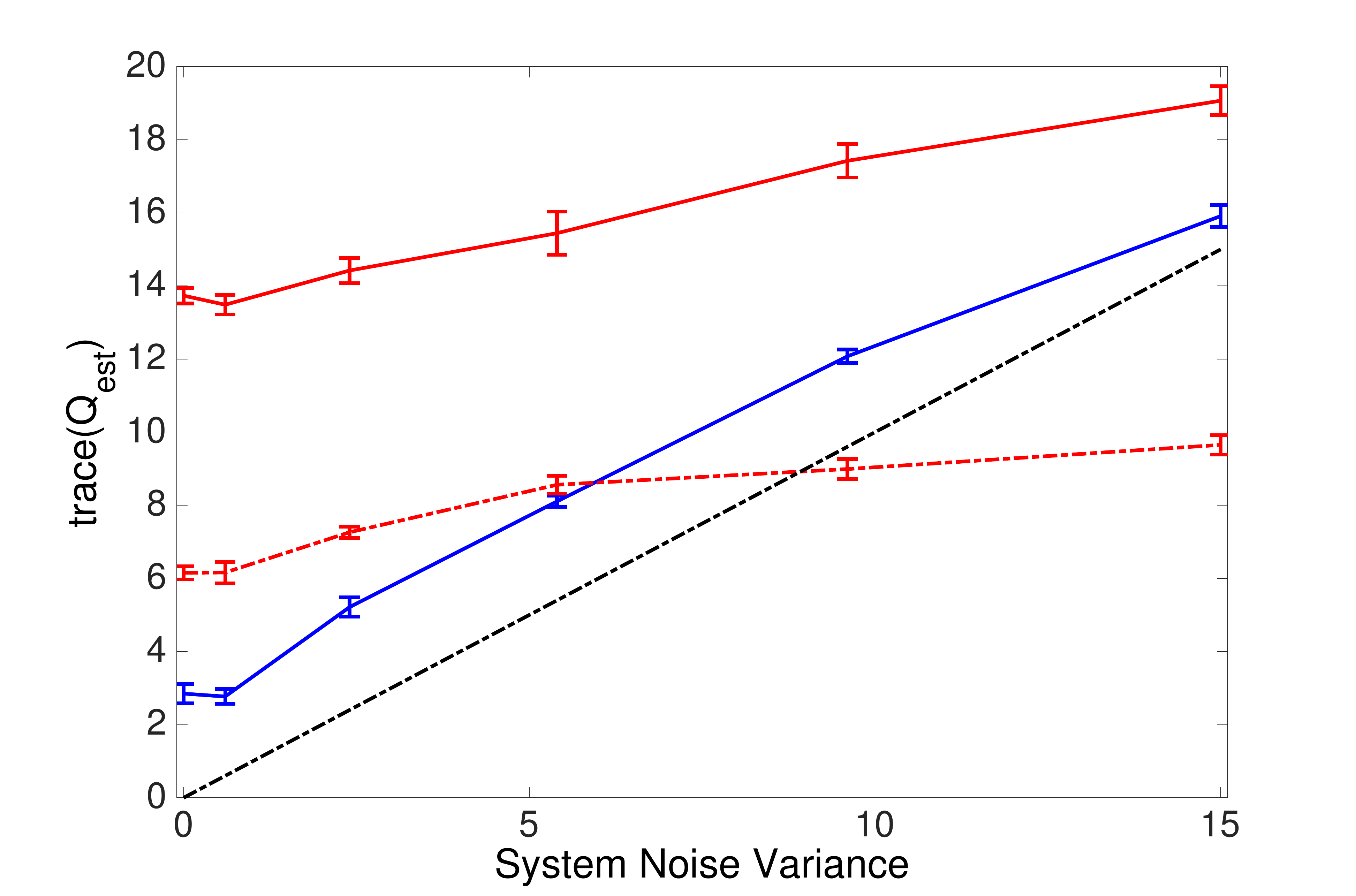}}
\subfigure[]{\includegraphics[width=.48\textwidth]{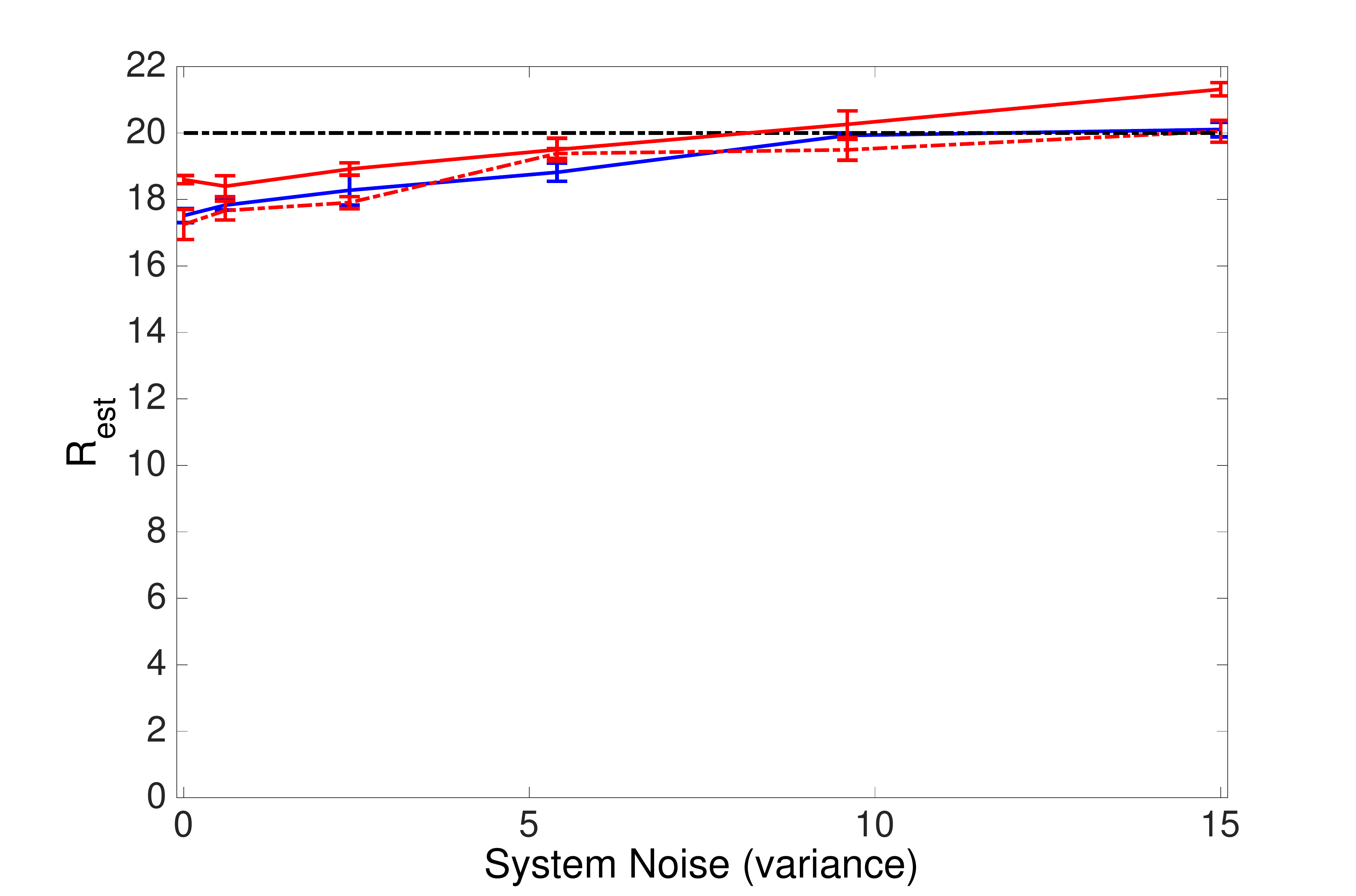}}
\end{center}
\caption{Adaptive estimation of the filter $Q$ and $R$ matrices. Parametric filter (solid blue curve), Kalman-Takens with 2 delays (solid red curve), Kalman-Takens with 4 delays (dotted red curve) (a) System noise estimate $Q_{est}$ (b) Observation noise estimate $R_{est}$}
\label{fig:5}       % Give a unique label
\end{figure}

%The estimates  are low-rank, noisy estimates of the covariance matrices $Q$ and $ R$ that will make the posterior estimate statistics from the filter consistent with the empirical statistics in the sense of (\ref{qrestimates}).  

To find stable estimates of $ Q$ and $ R$ we combine the noisy estimates $Q^e_{k-1}$ and $R^e_{k-1}$ (which are also low-rank by construction) using an exponential moving average
\begin{eqnarray}\label{qrupdate}
Q_{k} &=& Q_{k-1}+ (Q^e_{k-1}- Q_{k-1})/\tau \nonumber \\
R_{k} &=& R_{k-1} + (R^e_{k-1} - R_{k-1})/\tau. 
\end{eqnarray}
where $\tau$ is the window of the moving average. For further details on the estimation of $Q$ and $R$ we refer the reader to \cite{qr13,PRX}.

In Fig.~\ref{fig:5} we show the final estimates $\textup{trace}(Q_{est})$ and $R_{est}$ from applying the adaptive filter to the stochastic Lorenz system (\ref{e4}) with various system noise levels $3\xi^2$ (note that the factor $3$ gives the total variance of the three stochastic forcing variables which is compared to $\textup{trace}(Q_{est})$ which is also a total variance).  In Fig.~\ref{fig:5}(b) we show that all of the filter methods obtain reasonably accurate estimates of the true observation noise $R^o=20$, and the results are very robust to the amount of system noise.  In Fig.~\ref{fig:5}(a) we first note that the parametric filter with the perfect model obtains a good estimate (blue, solid curve) of the true system noise (black, dashed curve).  Secondly, we note that the trace of $Q_{est}$ for the Kalman-Takens filters both increase with increasing system noise, without distorting the estimate of the observation noise (as shown in Fig.~\ref{fig:5}(b)).  Notice that the Takens reconstruction corresponds to a highly nonlinear transformation of the state space, and that even the dimension of the state space changes.  Since this nonlinear transformation may stretch or contract the state space in a complicated way, it is not particularly surprising that the variance of the stochastic forcing in the delay-embedding space has a different magnitude than the original system noise variance.  The exact relationship between these two stochastic forcings is highly nontrivial, and the spatially homogeneous and uncorrelated system noise of (\ref{e4}) may easily become inhomogeneous and correlated in the reconstructed dynamics.  

%We interpret the moving average in (\ref{qrupdate}) as a moving average filter that stabilizes the noisy empirical estimates $Q_k$ and $R_k$.  The stochastic nature of the estimate of $Q_k$ can lead to excursions that fail to be symmetric and/or positive definite, leading to instability in the EnKF.  While the matrix $Q_k$ is not changed, the matrix used in the $k$-th step of the fter is a modified version of $Q_k$ which is forced to be symmetric and positive definite by taking $\tilde Q_k = (Q_k + Q_k^T)/2$ and then taking the max of the eigenvalues of $\tilde Q_k$ with zero.  Again, we emphasize that $\tilde Q_k$ is only used in the $k$-th filter step and no ad-hoc corrections are made to the matrix $Q_k$ which eventually stabilizes at a symmetric and positive definite matrix naturally via the moving average in (\ref{qrupdate}).  These ad-hoc corrections are only needed during the transient period of the estimation of $Q$, in particular when the trivial initialization $Q_1=0$ is used.  

\section{Filtering dynamical noise}
\label{sec:4}

\begin{figure}
\begin{center}
\subfigure[]{\includegraphics[width=.33\textwidth]{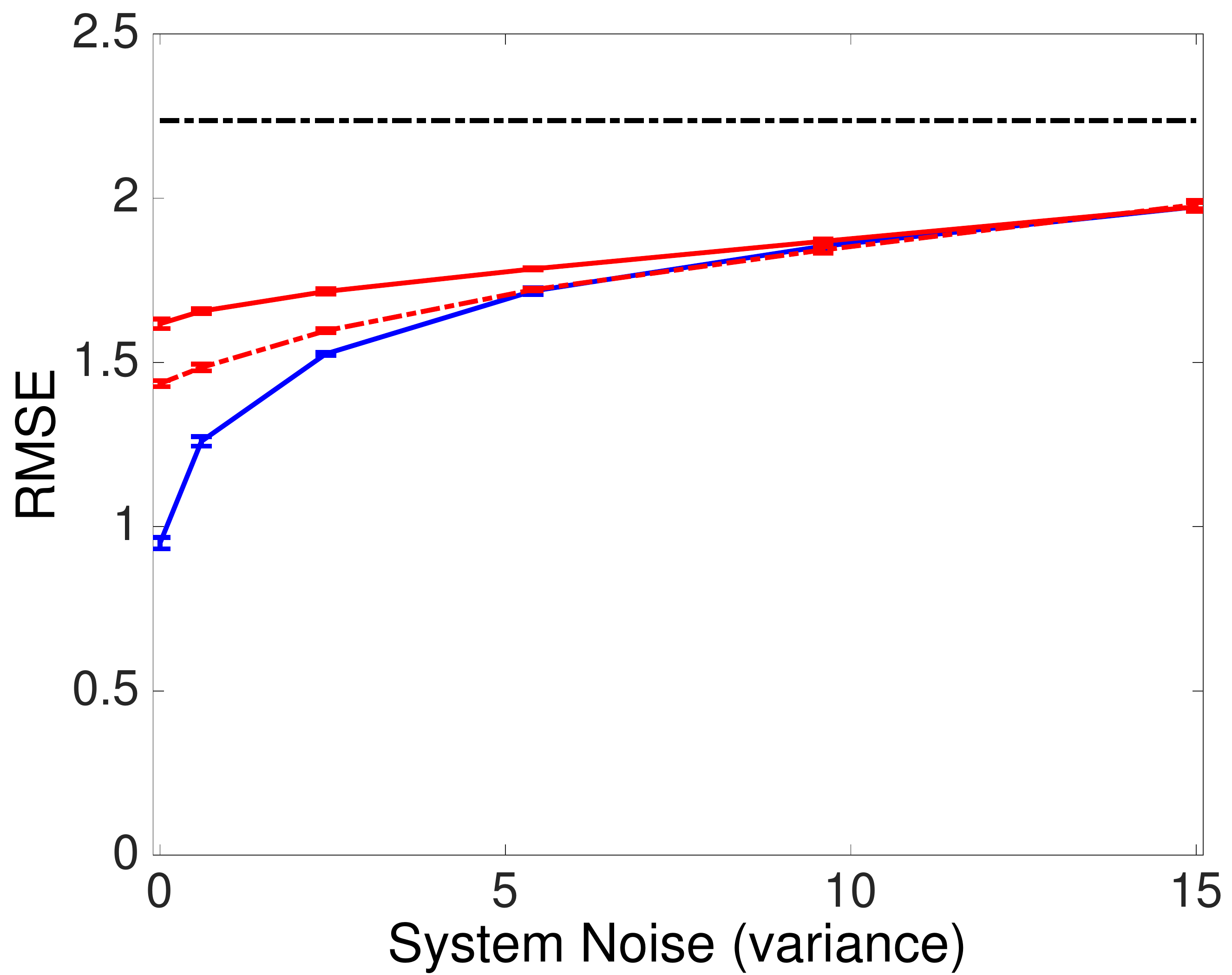}}
\subfigure[]{\includegraphics[width=.32\textwidth]{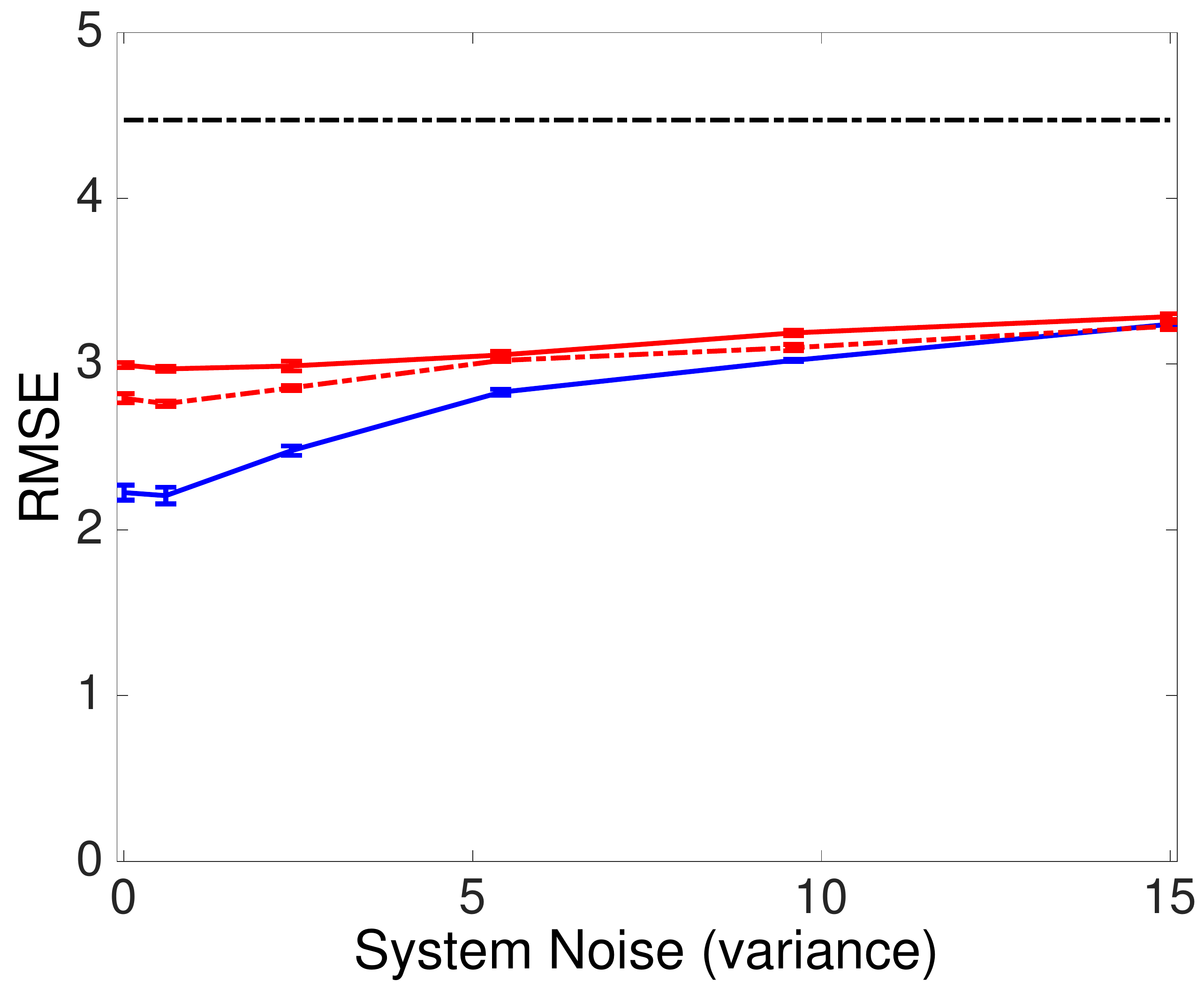}}
\subfigure[]{\includegraphics[width=.32\textwidth]{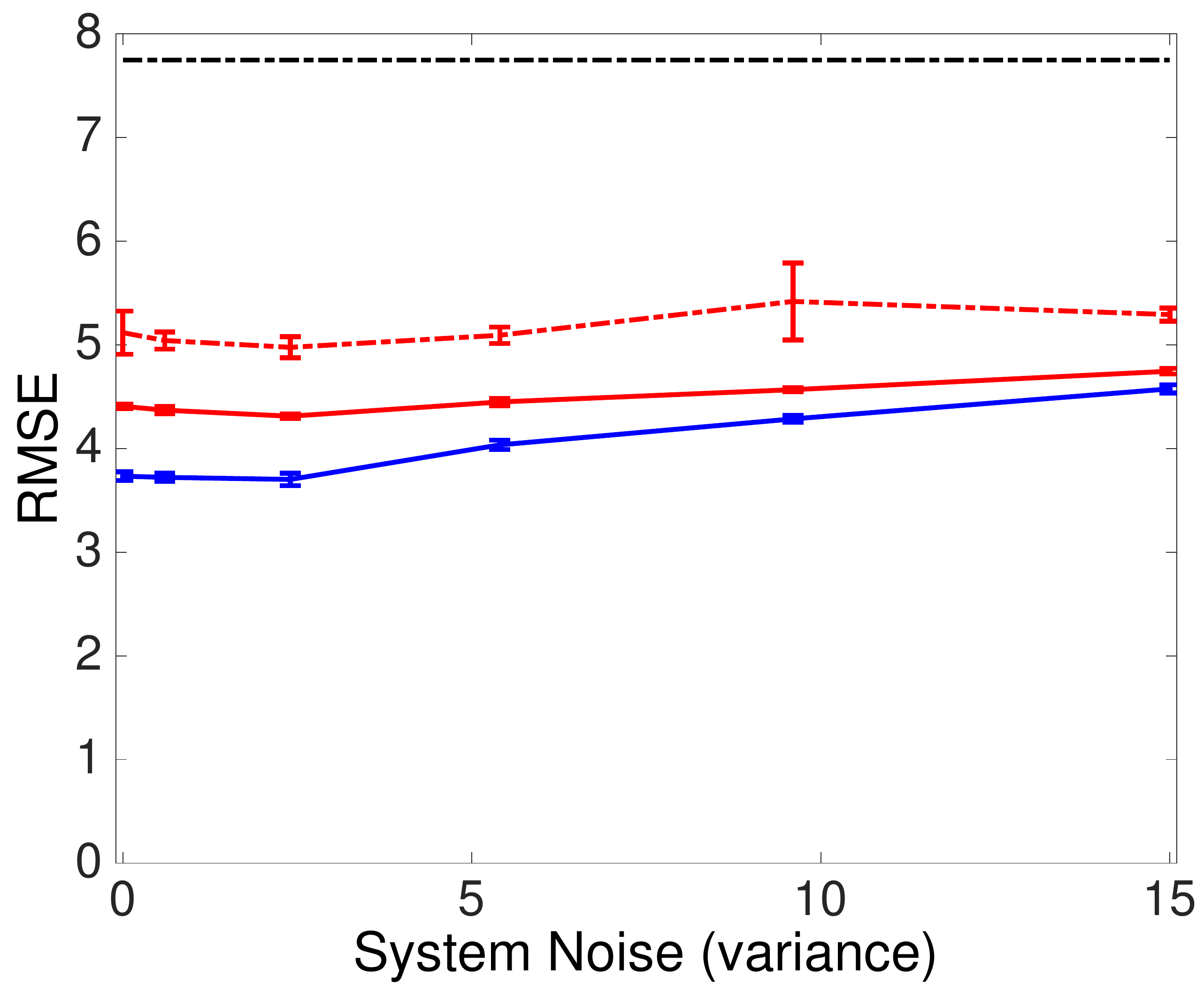}}
\end{center}
\caption{Reconstructing the noisy Lorenz-63 $x$ variable under increasing levels of dynamical noise.  Observations of stochastic signal perturbed by Gaussian observational noise with variance (a) 5, (b) 20 and (c) 60 resulting in signal error level (dotted black line). Parametric filter (solid blue line), Kalman-Takens filter with 2 delays (solid red line) and Kalman-Takens filter with 4 delays (dotted red line) reconstruction accuracy shown. Error bars denote standard error over 5 realizations. As the amount of dynamical noise increases, the Kalman-Takens filter has performance very similar to the parametric filter, which has access to the full model.}
\label{fig:3}       % Give a unique label
\end{figure}

The Kalman-Takens filter was introduced in \cite{PRX}, which considered deterministic dynamical systems with only observation noise.  
In section \ref{sec:2}, we showed that the filter is also robust to dynamical noise, and 
in this section we quantify this fact for stochastic systems such as (\ref{e4}) that include both system and observation noise.  In Fig.~\ref{fig:3} we compare the Kalman-Takens filters (red curves) to the parametric filter (blue curve) as a function of the system noise variance $3\xi^2$ for various levels of observation noise.  In Fig.~\ref{fig:3} the observation noise, which is the error between the observed signal and the true state, is denoted by the black dotted line in each subfigure.  Each filter obtains a dramatic denoising of the observations relative to the observation noise levels $R^o = 5, 20,$ and $60$ in Figs.~\ref{fig:3}(a,b,c) respectively.  In each case, for large observation noise the Kalman-Takens filter performance is comparable to the parametric filter, and only at very low system noise does the parametric filter have a slight advantage. This illustrates the robustness that the  Kalman-Takens filter has to model error since it makes no assumption on the form of the model, unlike the parametric filter which possesses the perfect model.  

We also explored the robustness of the Kalman-Takens filter to the number of delays $d$ used in the Takens delay-embedding.  For $d=2$ the embedding dimension is $d+1=3$, which is the minimum dimension needed to embed the attractor of the deterministic Lorenz-63 system.  However, the theoretical guarantee of the Takens embedding theory requires $d+1>2n$ where $n$ is the attractor dimension, which requires  an embedding dimension of $d+1=5$ (the attractor of the deterministic Lorenz-63 system is slightly larger than 2).  As shown in Fig.~\ref{fig:3}(a,b) the $d=2$ (red, solid curve) and $d=4$ (red, dotted curve) have similar performance except at very low system noise levels, where $d=4$ has a slight advantage.  Notice that when very little noise is present, the longer that two trajectories agree, the closer the corresponding states are, so for small noise, we expect long delays to help find better neighbors.  However, in the presence of system noise, this breaks down.  Two trajectories may be very close in the past but recent stochastic forcing may cause them to quickly diverge.  Moreover, as noise is added, it becomes increasingly difficult to distinguish states, and adding delays is merely adding more dimensions which could coincidentally agree without implying the states are similar.  This is shown in Fig.~\ref{fig:3}(c) where $d=2$ outperforms $d=4$ in the presence of large observation noise.  Similarly, in all of Fig.~\ref{fig:3} we note that as the system noise increases, the value of knowing the true model (\ref{e4}) becomes negligible as the $d=2$ Kalman-Takens performance is indistinguishable from the parametric filter.

\begin{figure}
\begin{center}
\subfigure[]{\includegraphics[width=.48\textwidth]{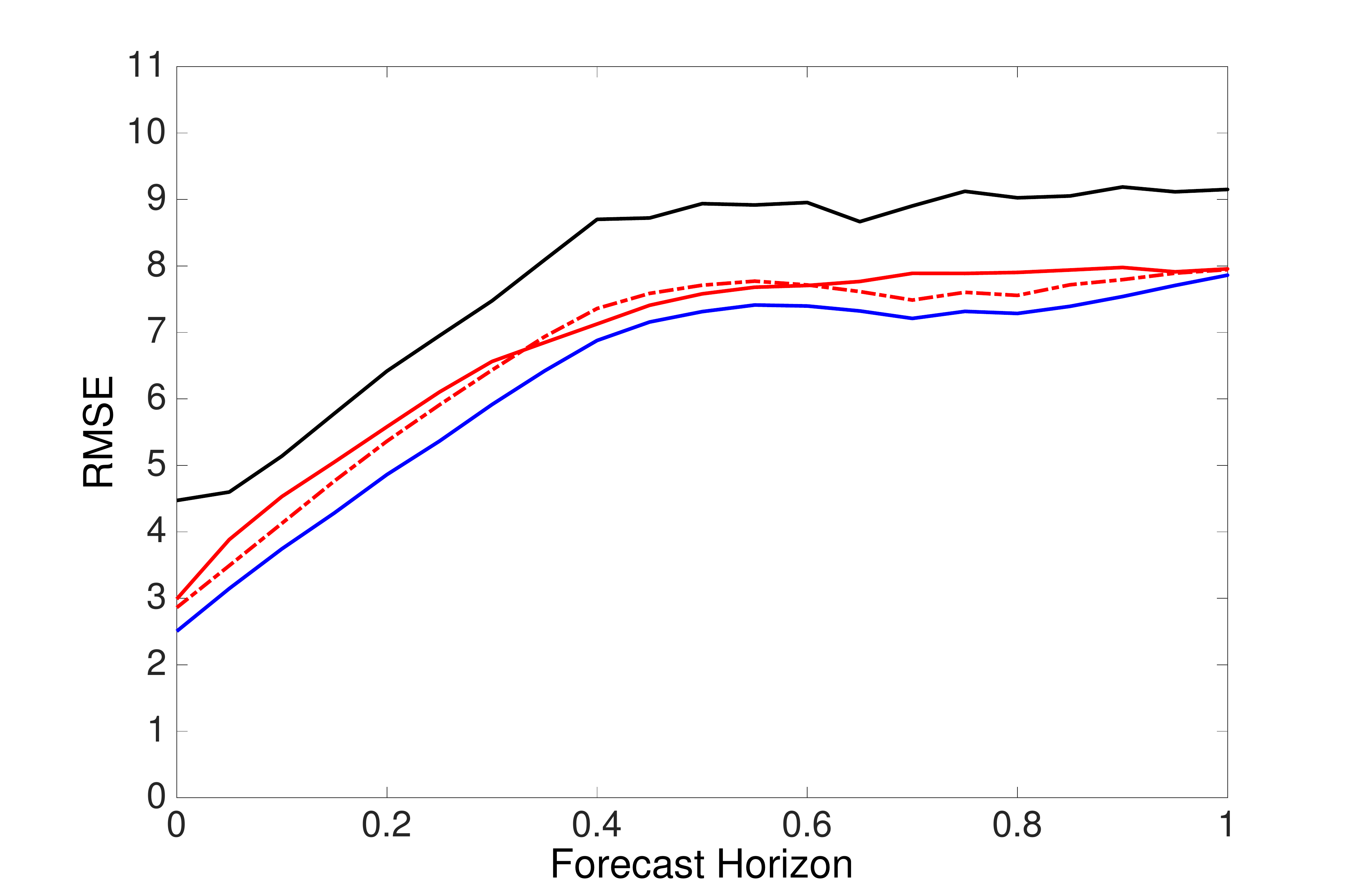}}
\subfigure[]{\includegraphics[width=.48\textwidth]{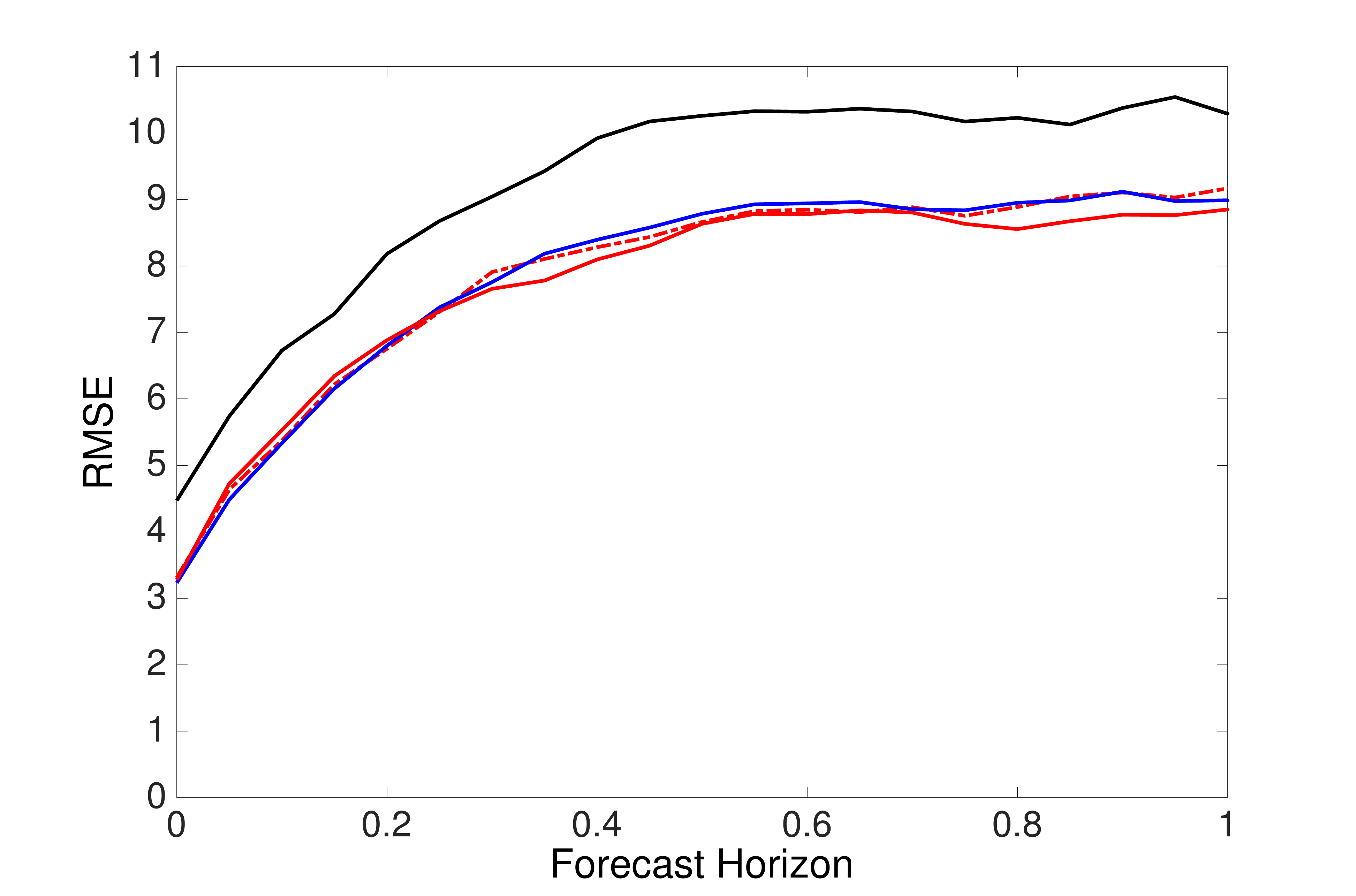}}
\end{center}
\caption{Forecast accuracy when predicting Lorenz-63 from 8000 noisy training data. (a) System influenced by Gaussian dynamical noise with variance of 0.8. (b) System influenced by Gaussian dynamical noise wtih variance of 5. Observations perturbed by Gaussian observational noise with variance of 20. Prediction results shown when the training data are not filtered (solid black curve), filtered by the parametric filter (solid blue curve), Kalman-Takens filter with 2 delays (solid red curve) and Kalman-Takens filter with 4 delays (dotted red curve) is used. Results averaged over 1000 realizations.}
\label{fig:4}       % Give a unique label
\end{figure}

Finally, we show in Fig.~\ref{fig:4} that the Kalman-Takens filter achieves forecast performance comparable to the perfect model.  The Kalman-Takens improves forecasting in two ways.  First, by running the Kalman-Takens filter on the historical training data, we significantly reduce the observation noise, which allows for better analog forecasting.  Second, the Kalman-Takens filter gives a good estimate of the current state, which leads to an improved forecast.  Ultimately, Takens based forecasting relies on finding good neighbors in delay-embedding space since these neighbors are interpolated to form the forecast.  Finding good neighbors requires both a good estimate of the current state, and historical data with small observation noise; and the Kalman-Takens filter improves both of these.  

As a baseline, we used the unfiltered delay-embedding and found neighbors in the raw delay-embedded historical data, and the RMSE of this forecast method is shown in Fig.~\ref{fig:4} (black solid curve) as a function of the forecast horizon.  The time zero forecast is simply the initial estimate, and for the black solid curve the RMSE at time zero corresponds to the observation noise level (since the initial state is simply the unfiltered current observation).  The lower RMSE of the filtered estimates (red and blue curves) at time zero in Fig.~\ref{fig:4} shows how the filter improves the initial estimate of the state.  Similarly, the long term forecast converges to the average of the historical training data (since we use an ensemble forecast which are uncorrelated in the long term).  The lower RMSE of the filtered forecasts (red curves) compared to the raw data (black curve) in Fig.~\ref{fig:4} shows that denoising the historical data gives an improved estimate of the long term statistics of the SDE.  Moreover, since the red curves are very close to the blue curve in Fig.~\ref{fig:4} we conclude that the Kalman-Takens is able to improve the historical data and estimate the current state with sufficient accuracy to match the parametric filter and forecast using the perfect model.

\section{Summary}
\label{sec:5}

Traditional data assimilation is predicated on the existence of an accurate model for the dynamics. In this article, we have shown that the Kalman-Takens filter provides an alternative when no model is available. Although it might be expected that dynamical noise would be an obstruction to the delay-coordinate embedding that is necessary to reconstruct the attractor, we find by numerical experiment that filtering and forecasting applications are not hampered significantly more than for the parametric filter.

This report is a feasibility study, and leaves open several interesting questions about how to optimally apply the algorithm. In particular, the role of the number of delays, neighborhood size, as well as their relation to the EnKF parameters are still not well understood, and are deserving of further investigation.

%\begin{table}
%\caption{Please write your table caption here.}
%\begin{tabular}{lll}
%\hline\noalign{\smallskip}
%\end{tabular}
%\end{table}
%

\end{document}